\documentclass{aa}
      
\usepackage{graphicx}

\newcommand{\be}{\begin{eqnarray}}
\newcommand{\ee}{\end{eqnarray}}
\newcommand {\COri} {\mbox{$\theta^1{\rm{C}}\:{\rm{Ori}}$}}
\newcommand {\Msun} {\mbox{M$_{\odot}$}}
\newcommand {\nbodypp}{\textsc{\mbox{nbody6\raise.4ex\hbox{\tiny++}}}}

\begin{document}

\title{The Fate of Discs around Massive Stars in Young Clusters}
\subtitle{}
\author{S. Pfalzner, C. Olczak, A. Eckart}
\offprints{}
\institute{I. Physikalisches Institut, University of Cologne, Germany}
\date{}
\authorrunning{Pfalzner et al.}
\titlerunning{Discs around Massive Stars}

\abstract
{}
{
The aim of this work is to understand whether there is a difference in the dispersion
of discs around stars in high-density young stellar clusters like the Orion Nebula Cluster 
(ONC) according to the mass of the star.
}{
Two types of simulations were combined ---  N-body simulations of 
the dynamics of the stars in the ONC and mass loss results from simulations 
of star-disc encounters, where the disc mass loss of all stars is determined
as a function of time.
}{
We find that in the Trapezium, the discs around high-mass stars are dispersed 
much more quickly and to a larger degree by their gravitational interaction than 
for intermediate-mass stars. This is consistent with the very recent 
observations of  IC 348, where a higher disc frequency was found around solar mass 
stars than for more massive stars, suggesting that this might be a general trend in 
large young stellar clusters.}
{}
\keywords{Accretion discs - young clusters - massive stars}

\maketitle

\section{Introduction}

Our current understanding is that planetary systems are formed from
the material of the star-surrounding protoplanetary discs and that the existence 
of a disc is a prerequisite for the formation of a planetary 
system. Not surprisingly there has been an increasing number of 
observational studies investigating the frequency of discs in young stellar 
clusters to address this question 
(\cite{hillenbrand:aj98,lada:aj00,haisch:apj01,kenyon:01}). The general trend seems
to be that the disc frequency decreases with the cluster age, but how quickly the disc 
dispersal process occurs is still not certain because due to observational 
constraints, the disc frequencies are often only known within a 
relatively wide margin. It might be that apart from the cluster age, 
factors like the presence of massive stars contribute to the 
disc dispersal rate.

In a recent study of IC 348 by \cite{lada:aj06} not only the disc 
frequency was investigated but also its dependence on the stellar 
mass. They found that the disc frequency is higher for intermediate-mass 
stars (47\%$\pm$12\%) than for massive stars (11\%$\pm$8\%) and low-mass 
stars (28\%$\pm$5\%).

At the moment several disc destruction mechanisms are under consideration,
among them photevaporation (\cite{storzer:99}) and gravitational interactions. 
Although favoured, photoevaporation mechanisms tend to have difficulties  
matching the observed disc destruction timescales (\cite{scally:mnras01}).

Here we will concentrate on modelling the effect of gravitational interactions.
Our simulations of the disc mass loss in the ONC show that high-mass stars 
lose their discs to a larger degree and more rapidly than intermediate-mass 
stars due to the gravitational interactions of the cluster stars. This effect 
could be a general tendency in all young clusters containing a sufficiently 
high number of massive stars.

\section{Model and Numerical technique}

We follow the idea of \cite{scally:mnras01}, combining a simulation of the
dynamics of the ONC to determine the interaction parameters of close encounters between
stars in the cluster with results from studies of isolated star-disc encounter
simulations. The main differences are slightly changed initial conditions of the cluster
and a more detailed treatment of the disc mass loss. The details of the numerics
can be found in \cite{olczak:06}. In contrast to the latter study the
emphasis in the current investigation is on the differences in disc mass loss for 
stars of low, intermediate and high mass. 

\subsection{Cluster simulation}

The dynamical model of the ONC presented here contains only stellar components without 
considering gas or the potential of the background molecular cloud OMC~1.
Cluster models were set up with a spherical density distribution $\rho(r)\propto r^{-2}$ and a Maxwell-Boltzmann 
velocity distribution. The masses were generated randomly according to the mass function 
given by \cite{kroupa:mnras93}
in a range $50 \Msun \ge M^* \ge 0.08 \Msun$, 
apart from \COri, which was directly assigned a mass of 50 \Msun \ and placed at the cluster centre.

\begin{figure}
\resizebox{\hsize}{!}{\includegraphics[angle=-90]{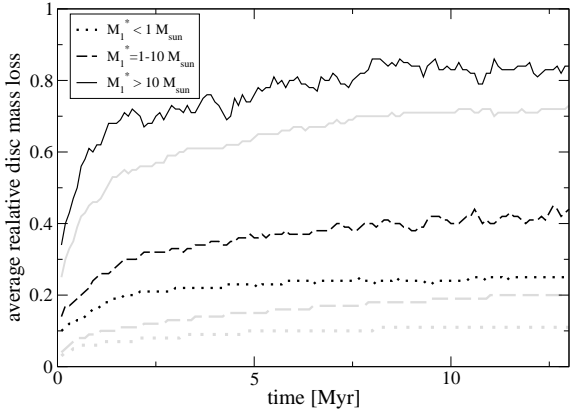}}
\caption{Temporal dependence of the average relative disc mass loss for stars of different mass
(black lines -- Trapezium, grey lines -- entire ONC).} 
\label{fig:zeit}
\end{figure}

In an encounter list the information of all perturbing events of each stellar 
disc was recorded during the course of the simulation, i.e. both masses, the relative 
velocity and the eccentricity.
The ONC was simulated for 13\,Myr -- the assumed lifetime of \COri.
The cluster simulations were performed with \nbodypp\ (\cite{spurzem:mnras02}). 
The quality of the dynamical models was determined by comparison to the observational data at 
1-2\,Myr, which marks the range of the mean ONC age. The quantities of 
interest were: number of stars, half-mass radius, number densities, velocity dispersion 
and projected density profile. Here we chose the ONC to be in virial equilibrium;
for more details of the selection process, see \cite{olczak:06}.

\subsection{Mass loss}

In the event of a close encounter between young cluster stars,  the discs surrounding them
can be severely disturbed and either partially or completely destroyed. From 
parameter studies for the disc mass loss in such encounters, \cite{pfalzner:aa05} 
determined a fit formula.
If one has to deal with penetrating encounters and high mass ratios 
$M_2^*/M_1^*$, like in the ONC, the following more complex Eq.~1 
\begin{eqnarray}
    \frac{\Delta M_{\rm{d}}}{M_{\rm{d}}}& = &  \left(\frac{M^*_2}{M^*_2+0.5M^*_1}\right)^{1.2} \log\left[2.8\left(\frac{r_{\rm{p}}}{r_{\rm{d}}}\right)^{0.1}\right]\nonumber \\
 \hspace{1cm} & \times &  \exp\left\{-\sqrt{\frac{M^*_1}{M^*_2+0.5M^*_1}}
    \left[\left(\frac{r_{\rm{p}}}{r_{\rm{d}}}\right)^{3/2}-0.5\right]\right\}~.
 \label{eqn:ImprovedFit}
\end{eqnarray}
is more suitable.
\begin{figure}
\resizebox{\hsize}{!}{\includegraphics[angle=-90]{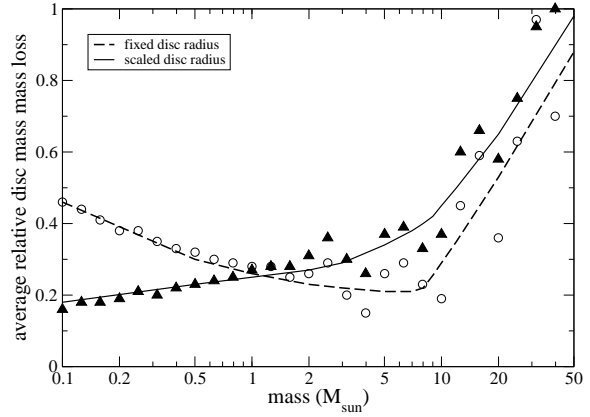}}
\caption{The average relative disc mass loss at 2 Myr for the Trapezium as a function of the
stellar mass. The triangles show the results assuming the disc radius to scale as in 
Eq.~\ref{eqn:rdisk} whereas the circles indicate the data for a constant disc radius(150 AU).}
\label{fig:disc_mass}
\end{figure}

However, a number of assumptions were made in these parameter studies:
only two-body encounters were considered, the discs were assumed to be of low mass 
(i.e., $M_{\rm{d}}/M^* \ll$ 0.1) and the surface density to have a $1/r$-dependence initially. 
In addition, Eq.~(\ref{eqn:ImprovedFit}) is for coplanar, prograde encounters only. According to studies 
on inclined and retrograde star-disc
encounters (\cite{clarke:mnras93,ostriker:apj94,heller:apj95,pfalzner:aa05}), 
these approximations  mean that Eq.~(\ref{eqn:ImprovedFit}) can only be 
interpreted as an upper limit of the disc mass loss.
However, the degree to which it overestimates the disc mass loss should be the same independent     
of the mass of the star, so that the quantitative numbers could be reduced, but this
should not change the qualitative results presented here.

As the cluster consists of a wide spectrum of stellar masses, the 
simulation results, valid for $M_1^*=1\,\Msun$, are generalized by scaling the disc radius 
according to
\begin{eqnarray}
    r_{\rm{d}}
    =150\,\mbox{AU}\sqrt{M_1^*[\Msun]}~,
    \label{eqn:rdisk}
\end{eqnarray}
which is equivalent to the assumption of a fixed force at the disc boundary. 
We will see in Section 3 what happens if we assume the disc radius 
to be uncorrelated to the mass of the star.

\section{Results}

\begin{figure}
\resizebox{\hsize}{!}{\includegraphics[angle=-90]{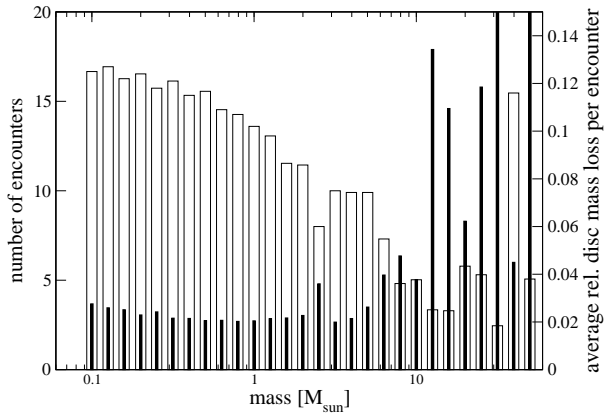}}
\caption{The average number of collisions that each star experienced up to 2 Myr is shown as black
bars as a function of the stellar mass for the Trapezium. For comparison the average relative disc 
mass loss per encounter is shown for the case of all discs initially being the same size.} 
\label{fig:no_enc}
\end{figure}

Applying Eq.~\ref{eqn:ImprovedFit} to the results
of the encounter tracking in 20 cluster simulations, we obtain the average disc mass loss
as a function of time. Looking at the temporal development of the disc mass of low-($M_1^*\leq$ 1.0 $\Msun$), intermediate-($M_1^*$=1.0--10$ \Msun$) and high-mass stars ($M_1^*>$ 10 $\Msun$) 
separately, Fig.~\ref{fig:zeit} shows that massive stars have a much higher disc mass loss 
than stars of lower mass, which manifests itself very early on in 
the cluster development. This is valid for the entire ONC but even more 
so for the central Trapezium cluster.

Taking a snapshot at 2 Myr (the ONC age is estimated to be 1-2 Myr and the
IC 348 age to be 2-3 Myr), the solid line in Fig.~\ref{fig:disc_mass} shows the average 
relative disc mass loss as a function of the mass of the star for the Trapezium region. 
At 2 Myr the disc mass loss is $\sim$ 30\% for intermediate-mass stars whereas for stars 
with $M^*$=20 $\Msun$ it is $\sim$ 60\% and for higher mass stars close to 100\%. 
This last value has to be taken with some care because of the poor mass loss statistics 
in this case.  

Scaling the disc size with the star mass (Eq.~\ref{eqn:rdisk}) seems intuitively right, 
but observational results do not give such a clear picture: Although \cite{vicente:aap05} 
derive a clear correlation between disc diameters and stellar masses using a sample of proplyds 
from \cite{luhman:apj00}, they see no indication for such a dependence in the data 
of \cite{hillenbrand:aj97}. However, as  \cite{vicente:aap05} pointed out, the present 
Trapezium is probably not in its primordial state and various disc destruction processes have 
most likely altered the disc sizes considerably.

If we assume that the disc size does not depend on the stellar mass as in Eq.~\ref{eqn:rdisk}
but is instead 150AU for all stars, this results 
in a somewhat smaller relative disc mass loss for the massive stars and an 
increase for the low-mass stars (see dashed line in Fig.~\ref{fig:disc_mass}).
Nevertheless, it still holds that the disc mass loss is considerably larger for massive stars than
for intermediate-mass stars. By contrast, for the low-mass stars the relative disc mass loss
is now higher than for intermediate-mass stars, leaving the stars with masses in the 
range 1-10 \Msun \ as the ones with the lowest disc mass loss in the cluster at 2Myr.
This is consistent with the observational results of \cite{lada:aj06}. 

To illustrate the underlying reason for this difference in relative disc mass loss, 
Fig.~\ref{fig:no_enc} shows the average number of encounters and average relative disc 
mass loss in a single collision as a function of the stellar mass. It can be seen that  
the number of collisions is nearly constant for 
low- and intermediate-mass stars but increases considerably for high-mass stars. 
So for low-mass stars
the mass-loss occurs through a few strong encounter events, whereas the disc of high-mass stars
is removed via a steady nibbling by many encounters with stars of lower mass.   

\begin{figure}
\resizebox{\hsize}{!}{\includegraphics[angle=-90]{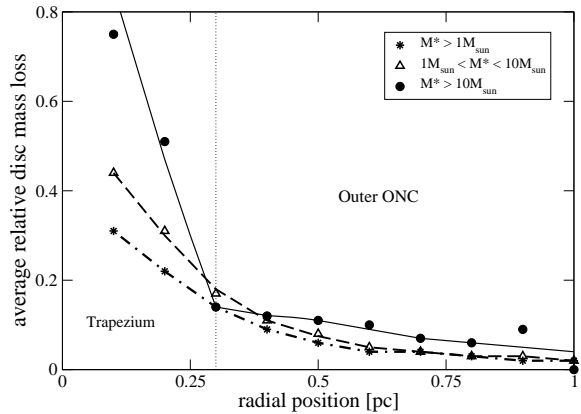}}
\caption{The average relative disc mass loss as a function of the radial distance from the
cluster center for low-mass, intermediate-mass and massive stars at 2 Myr. Here the disc size was assumed to be scaled. The dotted line indicates the extent of the Trapezium region.
} 
\label{fig:radius}
\end{figure}

Fig.~\ref{fig:radius} shows the average relative disc mass loss as a function of the
distance from the cluster center for low-, intermediate- and high-mass stars for the
entire ONC. For distances $r > $0.25 pc the massive stars do not exhibit higher disc mass 
loss than the low mass stars, but the number of massive stars at such radii is in any 
case small (3 stars with $M^* >$ 10 $\Msun$, 20 simulations), so that the statistics in 
this region is poor.

\begin{figure}
\resizebox{\hsize}{!}{\includegraphics[angle=-90]{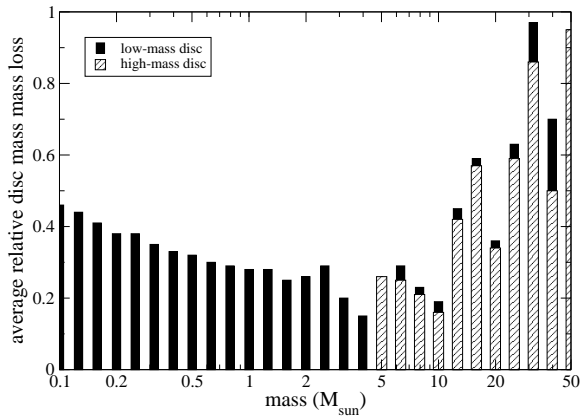}}
\caption{Average relative disc mass loss as a function of the stellar mass assuming constant 
disc radius. Comparison between the case where all discs are of low mass and
with the situation where all high-mass stars are surrounded by massive discs.} 
\label{fig:massive}
\end{figure}

During the last few years a number of massive stars surrounded by high-mass discs ($m_d>$0.1 $M^*$)
have been detected (\cite{zhang:05} and references therein). The question is how would the above 
results change if all massive stars had initially massive discs? As the interaction dynamics of
high-mass discs is much less understood than for low-mass discs, only an estimate can be given here.
To do so, we repeated the simulations with the assumption that all stars of M$^* >$ 5 $\Msun$ are 
surrounded by a disc of $m_d$=1 M$^*$ by simply assuming that the disc particles 
are all twice as strongly bound to their star, i.e. when Eq.~1 is applied, $M_2^*$ is
replaced by  $M_2^*+m_d$. This is obviously a strong simplification 
and more detailed investigations would be necessary in future.
However, Fig.~\ref{fig:massive} shows that 
although the stronger binding naturally leads to a smaller disc mass loss for the massive 
stars than before, they 
still lose a significantly higher proportion of their disc mass than intermediate-mass stars. 
In this case the described effect is thus only slightly weaker. 

\section{Summary and Discussion} 

In this article the relative disc mass loss induced by encounters between stars in a
cluster has been studied for the example of the ONC. The main result is that disc around
massive stars can be disrupted by close-by passages of stars with far greater efficiency than 
discs around intermediate-mass stars.
If we assume that the disc size scales with the stellar mass, the relative disc mass loss
is even lower for low-mass stars. However, if the disc size is independent of the stellar mass 
than low-mass stars have as well a higher relative disc mass loss then intermediate-mass stars
in accordance with \cite{lada:aj06}. In this
case we found the minimum disc mass loss for stars with 3\Msun$<M^*<$10\Msun. 

The reason for a higher disc mass loss for massive stars is twofold:
First, high-mass stars are found preferentially in the cluster core (due to dynamical mass 
segregation) where the stellar 
densities are higher and second the more massive stars suffer
larger disc mass loss than low-mass stars in the same region (as Fig.~4 shows), a result that 
can be traced (from Fig.~3) to the larger number of encounters suffered by these more massive 
stars.
The gravitational focussing of low-mass stellar orbits by massive
stars enhances the encounter rate between the massive star and other cluster 
members for any particular periastron separation.
So not only for the IC 348 as  observed by \cite{lada:aj06} and the ONC investigated here,  
but generally there should be more intermediate-mass stars than massive stars
surrounded by discs in the inner regions of high-density clusters, provided they
contain a sufficient number of massive stars.

In the inner cluster regions, discs around massive stars are much more rapidly destroyed 
than around intermediate-mass stars, so planetary systems should be less likely to develop around
massive stars. There the same should  hold for low-mass stars if the disc size is independent of 
the stellar mass.

The result that regardless of the stellar mass, fewer discs are destroyed at larger
distances from the center, does not necessarily mean a higher formation rate of planetary 
systems there. It actually depends on the dominating planet formation mechanism. If coagulation 
is the driving force this would hold. If on the other hand, planet formation predominantly
occurs via gravitational instabilities, then the higher encounter rate close to the cluster center
might even promote planet formation in the cluster core and possibly even dominate. However,
this effect has not been demonstarted to date.

In this paper we have concentrated on the destruction of discs caused by gravitational interactions 
of the cluster stars. Another disc destruction mechanism is 
photoevaporation (\cite{johnstone:98,storzer:99}). At the moment 
it is not entirely clear which mechanism dominates the disc destruction at a given cluster age.
Most likely gravitational interactions are more important during the first Myr and 
photoevaporation later on. However, as the disc mass loss is in both cases dependent on distance, 
qualitatively similar results - i.e. higher disc mass loss for the massive stars - should be 
obtained in the photoevaporation process as well.

\section*{Acknowledgments}
We want to thank R.Spurzem for his help by providing the Nbody6++ code for the cluster 
simulations. Part of the simulations were done on the 
JUMP Regatta of Research Center J\"ulich.

\end{document}